\documentclass[pra,twocolumn,preprintnumbers,amsmath,amssymb,amsfonts,superscriptaddress,showpacs,longbibliography]{revtex4-1}

\usepackage{graphicx}
\usepackage{amsmath}
\usepackage{graphics}
\usepackage{amssymb}
\usepackage{amsfonts,epsfig}
\usepackage{dsfont}
\usepackage{natbib}
\usepackage{enumitem}
\usepackage{physics}
\usepackage{color}
\usepackage[mathscr]{eucal}
\usepackage{ulem}

\newcommand{\REVIEW}{Szankowski_JPCM2017}		
\newcommand{\ACC}{Szankowski_PRA2018}			
\newcommand{\sinc}{\mathrm{sinc}}					
\newcommand{\attfc}[2]{\chi(#1|#2)}					
														
\begin{document}

\title{Transition between continuous and discrete spectra in dynamical-decoupling noise spectroscopy}
\author{Piotr Sza\'{n}kowski}\email{piotr.szankowski@ifpan.edu.pl}
\affiliation{Institute of Physics, Polish Academy of Sciences, al.~Lotnik{\'o}w 32/46, PL 02-668 Warsaw, Poland}

\begin{abstract}
Here, we explore the prospects of carrying out the single qubit spectroscopy of environmental noise when the resolution of the frequency filters produced by the dynamical decoupling pulse sequences can be set high enough to reveal the discrete structure of the noise spectral density. The standard form of spectroscopy is applicable when the filter is unable to resolve the discrete spectral lines and only coarse grained approximation of the spectrum is recorded in the qubit's decoherence rate. When the discrete structure becomes accessible, the qubit probe evolves in a qualitatively different manner, and the procedure for recovering spectral density has to be redesigned.
\end{abstract}
		
\maketitle

\section{Introduction}\label{sec:intro}

The mode of operating for methods of quantum metrology (such as dynamical decoupling spectroscopy) is to characterize the complex system of interest with the information acquired by a probe it is coupled to \cite{Degen_RMP2017}. Here we consider a case where the probe is realized with an efficiently and precisely controlled qubit (Q), that is used to scrutinize its environment (E) \cite{\REVIEW,Alvarez_PRL2011,Staudacher_SCI2013,Romach_PRL2015,Lovchinsky_SCI2016,Malinowski_PRL2017}. The inquired information is, thus, recorded and then retrieved from the course of the probe's evolution generated by the interactions governed by the coupling law $\hat H_\mathrm{int}$ that is specific to the particular physical implementation of the qubit. Therefore, the record made by the probe consists of valuable information on the internal dynamics of the environment (e.g., its initial state and free Hamiltonian $\hat H_\mathrm{E}$), that is inescapably distorted by the `lens' of qubit-environment coupling. The value of characterized internal dynamics is apparent because of its relevance for situations where the environment E becomes a part of any other experimental system. On the other hand, the probe-specific elements (such as the qubit-environment coupling law) are just that---specific to one particular setup. Hence, a perfect probe is one that allows for clean extraction of environment-only information or one where the distortions due to specific form of the coupling are negligible.

In the case of qubit-based dynamical decoupling spectroscopy \cite{\REVIEW,Viola_PRA1998,Viola_PRL1999,Ajoy_PRA2011} where the probe undergoes {\it pure dephasing} (i.e., when the free qubit Hamiltonian commutes with the interaction $[\hat H_\mathrm{Q},\hat H_\mathrm{int} ] = 0$) while being subjected to a periodic sequence of instantaneous $\pi$ pulses, the information is acquired in a form of {\it spectral density} that quantifies the distribution of frequencies belonging to the portion of environmental energy spectrum that is involved in its coupling to the probe. Such a probing scheme has been successfully implemented with various qubit designs: trapped ions \cite{Biercuk_Nature2009, Kotler_Nature2011}, superconducting circuits \cite{Bylander_NatPhys2011}, semiconductor quantum dots \cite{Medford_PRL2012,Dial_PRL2013}, ultracold atoms \cite{Almog_JoPB2011}, phosphorous donors in silicon \cite{Muhonen_NatNanTech2014}, and nitrogen-vacancy centers in diamond \cite{Staudacher_SCI2013,BarGill_NatComm2012,Romach_PRL2015}. 

The objective of the information gathering procedure designed for standard dynamical decoupling spectroscopy is to provide the means for the reconstruction of the course of spectral density. The most direct access to the required information is facilitated by the {\it Gaussian approximation} \cite{\REVIEW,\ACC} where spectral density (defined then as a Fourier transform of the auto-correlation of the temporal fluctuations of environmental degrees of freedom), together with the pulse sequence-induced frequency filter, fully describe the evolution of the qubit-probe. From this point we always assume that this approximation is in effect.

As stated previously, spectral density is an amalgamation of contributions from the environmental internal dynamics and the qubit-environment coupling law. Arguably the most direct piece of environment-only information that can be inferred from it is the central frequency positions of its pronounced peaks (the {\it spectral lines}) as well as their overall asymptotic behavior (e.g., the power-law tails or finite width, etc. \cite{\ACC,Abragam1961}) that reflects the structure of environmental energy spectrum. The intensity and precise shape of these peaks is of less relevance as they are determined, in large part, by the details of the qubit-environment coupling law. Here, we will demonstrate that it is possible to achieve a more robust quantification of the environmental spectrum through localization of spectral lines, whenever the resolution of the pulse sequence-induced frequency filters can be set beyond the limits normally used in the standard dynamical decoupling spectroscopy.

The outline of the paper is as follows. In Sec.~\ref{sec:overview}, we present an overview of the dynamical decoupling spectroscopy theory and we define the central question of this paper: How does the resolution of the pulse sequence-induced frequency filter affect the course of the qubit-probe evolution and its dependence on the settings of the sequence? Section~\ref{sec:sinc} provides an answer to this question. In Sec.~\ref{sec:disc_spectroscopy}, we discuss the consequences of the ability of high-resolution filters to distinguish discrete spectral lines, and their utility in spectroscopy. Then we formulate a scheme for the implementation of a variant of spectroscopy operating in this regime. In Sec.~\ref{sec:example}, we perform a numerical simulation that demonstrates its implementation in practice. We conclude the paper in Sec.~\ref{sec:end}.

\section{Theoretical overview}\label{sec:overview}
The state of the qubit-probe that undergoes pure dephasing due to coupling to the environment, as it is subjected to the periodic pulse sequence characterized by the {\it filter frequency} $\omega_p$, is of a following form \cite{\REVIEW,Anderson_JPSJ1954,Viola_PRA1998,Alvarez_PRL2011}
\begin{equation}
\hat\varrho_\mathrm{Q}(T) = \left(\begin{array}{cc}
\varrho_{{+}{+}} & \varrho_{{+}{-}}W(T|\omega_p) \\
\varrho_{{+}{-}}^* W^*(T|\omega_p) & \varrho_{{-}{-}}\\
\end{array}\right),
\end{equation}
here, the initial elements $\varrho_{ss'}=\langle{s}|\hat\varrho_\mathrm{Q}|{s'}\rangle$ ($s,s'=\pm$) are defined in respect to eigenstates of Pauli $z$ operator $\hat\sigma_z|\pm\rangle=\pm|\pm\rangle$. The filter frequency is related to the period of the applied pulse sequence $\omega_p = 2\pi/T_0$, and the duration of the evolution is chosen so that it encompasses a number of sequence periods $T=n T_0$, with integer $n$. The only evolving element of probe's state and, thus, the repository of information about the environment is the {\it coherence},
\begin{equation}
W(T|\omega_p) = e^{-\chi(T|\omega_p)},
\end{equation}
where the {\it attenuation function} $\attfc{T}{\omega_p}$ is real and $|\exp[-\attfc{T}{\omega_p}]|\leqslant 1$.

Within the Gaussian approximation, the attenuation function is given by the auto-correlation function of environmental fluctuations $C(t)$ and the time-domain filter function $f_{\omega_p}(t)$ that encapsulates the effects of the applied pulse sequence,
\begin{align}
\nonumber
\attfc{T}{\omega_p} &= \frac{1}{2}\int_0^Tdt_1dt_2 f_{\omega_p}(t_1)f_{\omega_p}(t_2)C(t_1-t_2)\\
\label{eq:attfc}
&= \frac{1}{2}\int_{-\infty}^\infty \frac{d\omega}{2\pi}|\tilde{f}_{\omega_p,T}(\omega)|^2S(\omega),
\end{align}
where $S(\omega)=\int_{-\infty}^\infty e^{-i\omega t}C(t)dt$ is the spectral density. The Fourier transform of the time-domain filter function is given by
\begin{align}
\nonumber
\tilde{f}_{\omega_p,T}(\omega) &=\int_{-\infty}^\infty e^{-i\omega t}\Theta(T-t)\Theta(t)f_{\omega_p}(t)dt\\
&= \sum_m c_{m}h_T(\omega-m\omega_p),
\end{align}
where the function.
\begin{align}
\label{eq:passband}
h_T(\omega) &= \int_{-\infty}^\infty \!\!\!e^{-i\omega t}\Theta(T-t)\Theta(t)dt
= T e^{-i \frac{T\omega}{2}}\sinc\left(\frac{T\omega}2\right)
\end{align}
is the shape of the filter's frequency passbands with the width given by the inverse of the duration, and the weights,
\begin{equation}
\label{eq:c}
c_{m} = \frac{1}{T_0}\int_0^{T_0}e^{-i m\omega_p t}f_{\omega_p}(t)dt
\end{equation}
are the Fourier series coefficients of the filter function defined with respect to its single period $T_0$ (see Appendix \ref{sec:filter} for a detailed derivation). Note that only the width of the passbands depends on the duration---the direct consequence of the periodicity of pulse sequences. This feature allows us to treat $T$ as an independent parameter and to manipulate its length without affecting any other characteristic of the frequency filter (which includes the filter frequency $\omega_p$ and coefficients $c_{m}$). This ability to manipulate $T$ independently of other parameters is crucial in our further considerations, and so, the periodicity of the pulse sequence is a necessary prerequisite. In the case of pulse sequences that are inherently periodic, e.g., where pulses are applied in equal intervals, the period is easily defined (if the interpulse interval is $\tau_p$, the period is simply $T_0 = 2\tau_p$). However, if the sequence is not periodic in such a way, e.g., the pulse timings have been chosen at random within a fixed evolution duration, then the periodicity can still be restored by treating the sequence as a base block of pulses that can be repeated an arbitrary number of times. The period of sequences constructed in such a way equals the duration of the original sequence, and the duration $T$ is controlled by choosing the number of base block repetitions $n$. Contrast this form of duration manipulation through period repetitions with other forms considered in the context of dynamical-decoupling-based noise analysis. For example, in Ref.~\cite{Pasini_PRA2010}, the considered sequences consist of fixed number of pulses $n_0$ with varying interpulse interval $\tau_p$. Obviously, by changing $\tau_p$, one also changes the duration as $T \propto \tau_p$. However, this also changes the period of the sequence, and in consequence, the filter frequency $\omega_p$ and the Fourier coefficients $c_m$. Thus, the duration can no longer be considered as an independent parameter.

Previous analysis of the behavior of the attenuation function as a function of duration \cite{\ACC} assumed implicitly that the spectral density $S(\omega)$ is a continuous function. Under this assumption, it has been shown that $\chi(T|\omega_p)$ is a combination of three terms, each adhering to a different $T$-{\it scaling law}, i.e., exhibiting a characteristic type of dependence on $T$, when $T$ is an independent parameter:  
\begin{equation}
\chi(T|\omega_p) = \frac{T}{2}\sum_m|c_m|^2 S(m\omega_p)+\Delta\chi_0+\Delta\chi_T.
\end{equation}
The so-called {\it spectroscopic formula} that scales linearly with $T$ and its correction terms $\Delta\chi_0$ that is independent of $T$ and $\Delta\chi_T$ that decays, at least, as fast as the correlation function $C(T)$. This allows for setting up a data-acquisition scheme where by exploiting the distinct parametric dependence on the duration, one is able to extract the spectroscopic formula from among other terms, thus, enabling an accurate spectrum reconstruction.

However, the continuous functions do not exhaust all possible forms of spectral densities. The key example is the case of quantum noise where the qubit-environment coupling law is given by the interaction Hamiltonian $\hat H_\mathrm{int}=\hat V\hat\sigma_z/2$, with $\hat V$ operating in environmental Hilbert subspace. When the free Hamiltonian of the environment possesses a discrete spectrum $\hat H_\mathrm{E} = \sum_{i} E_i |i\rangle\langle i |$ and its initial state $\hat\varrho_\mathrm{E}$ is stationary $[\hat H_\mathrm{E},\hat\varrho_\mathrm{E}]=0$, the auto-correlation function has the following form \cite{\REVIEW},
\begin{align}
\nonumber
C(t) &= \frac{1}{2}\mathrm{Tr}\left( \hat\varrho_\mathrm{E}\,\{ e^{i t\hat H_\mathrm{E}}\hat V e^{-i t\hat H_\mathrm{E}}\,,\,\hat V \}\right)\\
&= \sum_{i,j}p_i |V_{ij}|^2\cos[ (E_i-E_j)t]\,.
\end{align}
Here, $\hat\varrho_\mathrm{E}= \sum_{i} p_i |i\rangle\langle i|$ (with $p_i\geqslant 0$ and $\sum_{i}p_i=1$), $\{\hat A,\hat B\}=\hat A\hat B+\hat B\hat A$, and $V_{ij} = \langle i |\hat V|j\rangle$. The noise spectral density is then obtained by transforming this expression to the frequency domain, which leads to what we will refer to as the {\it discrete spectral density}:
\begin{align}
&S(\omega)\! =\! \pi\!\sum_{i,j} p_i |V_{ij}|^2[\delta(\omega-E_i+E_j) + \delta(\omega+E_i-E_j)]\,.
\end{align}
When such a form of spectral density is substituted to the general expression for attenuation function \eqref{eq:attfc}, one obtains the following result:
\begin{align}
\nonumber
\attfc{T}{\omega_p} =&\frac{1}{2}\sum_{\omega_k}I(\omega_k) \sum_{m,m'}c_{m}c_{m'}^*\\
\label{eq:disc_chi}
&\times h_T(\omega_k-m\omega_p)h_T^*(\omega_k-m'\omega_p),
\end{align}
where we defined the the {\it intensity} of the spectral line located at frequency $\omega_k$,
\begin{align}
\label{eq:matrix_elem}
I(\omega_k) &=\!\!\sum_{i,j:|E_i-E_j|=|\omega_k|}\!\!(p_i+p_j) |V_{ij}|^2 = I(-\omega_k).
\end{align}

In the right conditions, when the discrete line distribution is dense and the intensities vary relatively smoothly from line to line, the discrete spectrum can become indistinguishable from its continuous or {\it coarse-grained} approximation. Whether the discrete aspect of the spectrum is apparent depends on how the line density (quantified by a typical separation between neighboring line positions $\omega_k$) compares to the width of the passband functions of the frequency filter $T^{-1}$.  Intuitively, if the passbands are much wider than the typical distance between lines, i.e., $T\overline{\Delta\omega}\ll 1$, the discrete nature of the spectral density should not be detectable, and the sum over frequencies $\omega_k$ become well approximated by an integral with continuous, coarse-grained spectral density. In such an event, the structure of the attenuation function and the $T$-scaling laws described in Ref.~\cite{\ACC} would be observed here as well. The quantitative conditions for applicability of this kind of approximation are explored in more detail in Appendix~\ref{sec:cont_limit}. When the coarse-grained picture does not hold anymore because the passband functions are not wide enough to overlap with more than one spectral line, i.e., $T> 2\pi/\overline{\Delta\omega}$, the frequency-domain filter produced by the pulse sequence should be able to resolve discrete distribution of $\omega_k$. Of course, in such a case, the previously obtained results for continuous spectral densities would no longer be correct, and the $T$-scaling laws of the attenuation function, as well as their utility for spectroscopy ought to be reconsidered. 

Here, we will focus our attention on spectral densities that have the ability to exhibit this kind of transition from coarse-grained continuous function to discrete line distribution when the filter resolution is enhanced. According to what we said above (see also Appendix \ref{sec:cont_limit}), such spectra comprise a large number of relatively densely packed and comparably intense spectral lines. This means that we are {\it not} considering here the cases of sparse discrete line distributions \cite{Childress_Science06,Zhao_NN12,Taminiau_PRL2012,Krzywda_PRA2017,Hernandez_PRB2018} (characteristic for, e.g., small environments), nor a sparse discrete distribution superposed with a continuous background \cite{Zhao_PRL2011,Kolkowitz_PRL2012}, encountered, e.g., when the coupling between the probe and a small fraction of larger environment is particularly strong.

\section{$\boldsymbol{T}$-scaling laws for discrete spectral densities}\label{sec:sinc}

\begin{figure}[tb]
\centering
\includegraphics{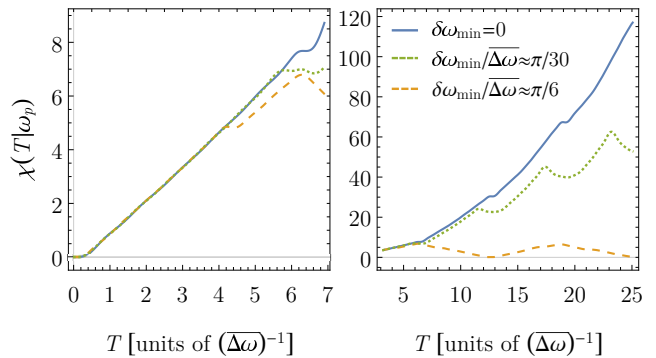}
\caption{The duration scaling of attenuation function for discrete spectral density. For the purpose of more transparent presentation, we treated here $T$ as a continuous parameter, instead of restricting its values to integer multiples of pulse sequence period $T_0$. The figure on the left depicts the duration scale---the coarse-grained regime---where the width of the passbands of the pulse sequence-induced frequency filter (proportional to $T^{-1}$) is large enough to encompass many spectral lines so that the spectral density is well approximated by continuous function. The resultant linear scaling of $\chi(T|\omega_p)$ is observed for all values of filter frequency $\omega_p$. The figure on the right shows types of $T$-scaling for duration beyond coarse-grained regime where the passbands are too narrow to overlap with more than one spectral line (i.e., $T>2\pi/\overline{\Delta\omega}$). The observed scaling laws depend on the value of $\omega_p$ via its detuning from the nearest spectral line position quantified by $\delta\omega_\mathrm{min}$ defined in Eq.~\eqref{eq:detunin}. The attenuation function illustrated here was calculated according to Eq.~\eqref{eq:attfc} with a setup identical to one used in Sec.~\ref{sec:example}.
}
\label{fig:transitions}
\end{figure}
Figure~\ref{fig:transitions} showcases the typical behavior of $\attfc{T}{\omega_p}$ as a function of duration $T$ for discrete spectral density: Initially, we can see a linear scaling, characteristic for spectroscopic formula, which signifies that the continuous spectral density approximation holds firmly for this duration scale---we will refer to this section as the {\it coarse-grained regime}. As the duration increases to post-coarse-grained regime where the continuous approximation breaks down, we observe the emergence of qualitatively different types of scaling laws that depend on the setting of filter frequency $\omega_p$. More precisely, the parameter that controls the type of scaling law in post-coarse-grained regime is the {\it minimal detuning} defined as
\begin{equation}
\label{eq:detunin}
\delta\omega_\mathrm{min}\equiv \min_{\omega_k, m}\left|\omega_k-m\omega_p\right|.
\end{equation}
It describes the shortest distance between the filter frequencies (given by multiples of $\omega_p$ present in Fourier series decomposition of filter function) and the positions of spectral lines $\omega_k$. Here we wish to reiterate that, although vital after the transition from coarse-grained regime, the value of $\delta\omega_\mathrm{min}$ has no bearing on the scaling laws encountered when the continuous spectrum approximation is still in effect.

The type of post-coarse-grained regime scaling laws and their dependence on value of minimal detuning can be qualitatively explained with the help of simplified model where one neglects the complications introduced by the side passbands in the exact attenuation function \eqref{eq:disc_chi},
\begin{align}
\nonumber
&\chi(T|\omega_p)\approx 2|c_{1}|^2\sum_{\omega_k}I(\omega_k)|h_T(\omega_k-\omega_p)|^2\\
\label{eq:chi_model}
&\propto I(\omega_0)\frac{4\sin^2\left(\frac{T\delta\omega_\mathrm{min}}2\right)}{\delta\omega_\mathrm{min}^2}
	+\sum_{\omega_k\neq \omega_0}\!\!\! I(\omega_k)\frac{4\sin^2\left(\frac{T\delta\omega_k}2\right)}{\delta\omega_k^2},
\end{align}
where $\delta\omega_k = |\omega_k-\omega_p|$ and $\omega_0$ denotes the spectral line position for which the detuning is minimal $\delta\omega_0 = \delta\omega_\mathrm{min}$.

In the special case of $\delta\omega_\mathrm{min}=0$, i.e., when the filter frequency matches {\it exactly} one of the discrete spectral lines, the resonant passband function becomes $4\sin^2(T\delta\omega_\mathrm{min}/2)/\delta\omega_\mathrm{min}^2 = T^2$, so that the attenuation function scales with $T$ as the second degree polynomial. For non zero minimal detuning, the polynomial scaling is sustained only on the duration scale for which it is appropriate to approximate the sine function with a few lowest-order terms of its power series, the extent of which can be estimated as $T\delta\omega_\mathrm{min}/2\ll\pi/2\Rightarrow T\ll \pi/\delta\omega_\mathrm{min}$. For longer duration, the oscillatory behavior of the sine takes over, and the scaling turns into harmonic oscillations with the period determined by the detuning $T_\mathrm{osc}=4\pi/\delta\omega_\mathrm{min}$.

The contributions from the remaining, off-resonance spectral lines are always in the harmonic-oscillation mode of scaling since, for them, the polynomial approximation of $4\sin^2(T\delta\omega_k/2)/\delta\omega_k^2$ breaks down simultaneously with the transition from the coarse-grained regime $T>2\pi/\overline{\Delta\omega}>2\pi/\delta\omega_k\approx 2\pi/(\delta\omega_\mathrm{min}+k\overline{\Delta\omega})$. Moreover, the resultant oscillation periods are shorter than $T_\mathrm{osc}$ because the corresponding detunings are, by definition, larger than $\delta\omega_\mathrm{min}$. For the same reason, the amplitudes of these oscillations are also much smaller due to inverse proportionality to $\delta\omega_k^2$ of each term. Hence, the off-resonance lines add a beating patter over the $T$ scaling due to the in-resonance spectral line.

The explanation of relations between detuning and the emergent scaling laws described above can still be applied to the exact form of the attenuation function (when the side passbands are not neglected). Similar to the simplified model, the scaling law is determined by the minimally detuned term proportional to $|h_T(\delta\omega_\mathrm{min})|^2$. A new element introduced by the side passbands is more complex background beating patterns due to the addition of mixed terms of form $h_T(\omega_k-m\omega_p)h_T^*(\omega_k-m'\omega_p)$ where, at most, only one of the passband functions can be in resonance. More importantly, it is vital to note that, according to definition \eqref{eq:detunin} of minimal detuning $\delta\omega_\mathrm{min}$, there is a possibility that the resonance occurs between spectral line and one of the side filter frequencies $m\omega_p$ with $m>1$. This observation is crucial for our upcoming discussion on possible designs of the spectroscopy procedure that operates in post-coarse-grained regime, where the side passbands of the filter have to be taken into account.

\section{Spectroscopy of discrete spectral density}\label{sec:disc_spectroscopy}
When the polynomial scaling is observed in measured attenuation functions one could, in principle, estimate the value of polynomial coefficients (e.g., by performing an appropriate fit to data points in an analogous fashion as is performed for coarse-grained regime spectroscopy \cite{\ACC}) in order to recover the value of line intensity $I(\omega_k)$. However, even if such estimates could be performed accurately, found intensities are arguably less informative than the straightforward information on the location of the spectral line itself $\omega_k$. According to Eq.~\eqref{eq:matrix_elem}, $I(\omega_k)$ contains information on the initial state of the environment in a form of probabilities $p_i$ and on the particular qubit-environment coupling law in a form of matrix elements $|V_{ij}|^2$. However, all these quantities are entangled in such a way that is not clear that it is even possible to unravel them. On the other hand, the spectral line positions provide a direct insight into the structure of the environmental Hamiltonian $\hat H_E$---the characterization of its internal dynamics. Therefore, whenever it is possible to reach beyond the coarse-grained regime, we propose to shift the goal of the spectroscopy from the traditional reconstruction of the shape of spectral density (which is a natural objective in the coarse-grained regime) to the reconstruction of the discrete distribution of spectral lines. Of course, the information on localization of the lines can always be supplemented with the results of the reconstruction of the coarse-grained spectral density in order to obtain an estimation of line intensities.

The objective of this proposed type of post-coarse-grained regime spectroscopy is to determine the value of detuning $\delta\omega_\mathrm{min}$ for given $\omega_p$---the smaller the detuning, the closer the filter frequency matches the frequency of one of the spectral lines, thus, reveling its position. On the other hand, the value of $\delta\omega_\mathrm{min}$ also impacts the $T$-scaling laws of qubit's attenuation function: the monotonic growth of polynomial scaling when the filter frequency is in-resonance and oscillatory behavior when $\omega_p$ is detuned from any intense spectral line. Since these types of $T$ dependence are so dramatically different, it is enough to observe the course of evolution of coherence $W(T|\omega_p)$ in order to determine which mode of scaling is currently active. Below, we define and discuss the simplest procedure that adheres to such a design principle.

\begin{figure}[tb]
\centering
\includegraphics{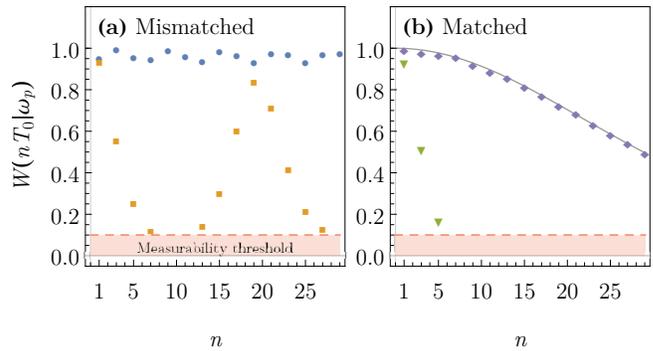}
\caption{(a) Examples of the plot of coherence evolution classified in the {\it mismatched} category. Fast oscillations with small amplitude (blue circles) and the revival pattern (orange squares) correspond to minimal detuning $\delta\omega_\mathrm{min}$ that is, respectively, large ($\delta\omega_\mathrm{min}\gg 2\pi/T_\mathrm{max}\propto n_\mathrm{max}^{-1}$) and of intermediate size ($\delta\omega_\mathrm{min}\sim 2\pi/T_\mathrm{max}$). (b) Plots of coherence evolution classified in the {\it matched} category. The green triangles depict the case of resonance with intense spectral line ($\delta\omega_\mathrm{min}=|\omega_0-\omega_p|= 0$ and $T_\mathrm{max}\sqrt{I(\omega_0)}\approx 12$) where the coherence decays monotonically below detection threshold and no revivals are observed on the recorded duration scale. In the case of resonance with a line of low intensity ($\delta\omega_\mathrm{min}=0$, $T_\mathrm{max}\sqrt{I(\omega_0)}\approx 1$), the coherence decays to finite value (purple diamonds). The gray line presents the best fit to measured data points of polynomial scaling of attenuation function, i.e., $W(T) = \exp[- u T^2/4+u \delta\omega^2 T^4/48]$ (with $u$ and $\delta\omega$ being the fit parameters). According to the obtained fit, the estimated detuning is one order of magnitude smaller than the plot resolution, $\delta\omega/(2\pi/T_\mathrm{max})\approx 0.3$. The measured data were simulated with the same setup as in Sec.~\ref{sec:example}.}
\label{fig:coherence}
\end{figure}

In the first step, we choose filter frequency $\omega_p$ and record the values of coherence in a series of measurements with increasingly longer duration $\{ |W(T_1|\omega_p)|,|W(T_2|\omega_p)|\ldots, |W(T_\mathrm{max}|\omega_p)|\}$, where $T_i = n_i T_0 = n_i 2\pi/\omega_p$, $T_\mathrm{max} = n_\mathrm{max} 2\pi/\omega_p$, and $n_1<n_2<\ldots<n_\mathrm{max}$. Of course, the shortest duration in a series $T_1$ must be long enough to make a transition to the post-coarse-grained regime. Next, we plot the gathered data set on a graph in order to inspect the course of coherence evolution. The obtained plot can be classified into two categories as illustrated in Fig.~\ref{fig:coherence}: (i) The coherence exhibits oscillatory behavior that is manifested as rapid low-amplitude oscillations on one end of the spectrum, and on the other end, we have a {\it revival} pattern where coherence decays to a deep local minimum and rises afterwards towards a high local maximum. The presence of oscillations indicates that the minimal detuning is relatively large ($\delta\omega_\mathrm{min}\gtrsim 2\pi/T_\mathrm{max}$ since the recording captured, at least, a half-period of the slowest oscillation). In such a case, we take a stance that $\omega_p$ is {\it mismatched} with positions of all the spectral lines with significant intensity. (ii) The coherence decays monotonically. When the decay brings the coherence to zero (or, at least, below a measurable threshold), and there are no visible revivals on the observed duration scale, we can conclude that $\delta\omega_\mathrm{min}\ll 2\pi/T_\mathrm{max}$. If the coherence remains finite, one can estimate the value of $\delta\omega_\mathrm{min}$ more precisely by looking for the best fit of polynomial decay to the data points $W(T)=\exp[-u T^2/4+u \delta\omega^2 T^4/48 +o(T^6)]$, with fit parameters $u$ and $\delta\omega$; the latter parameter can be considered as an estimate of minimal detuning. Whatever the case, the lack of oscillatory behavior and visible polynomial $T$-scaling means that the detuning is small and so, either the filter frequency $\omega_p$ or one of the side frequencies $m\omega_p$ ($m>1$), has {\it matched} the position of the spectral line. 

By examining the course of qubit's evolution in this manner for a wide range of filter frequencies, we can effectively scan the frequency domain and search for those choices of $\omega_p$ that produced plots of $W(T|\omega_p)$ categorized as matching. However, finding matching $\omega_p$'s does not conclude the procedure yet as, at this stage, we cannot say which side frequency $m\omega_p$ is, in fact, in-resonance for a given choice of filter frequency. In order to determine $m$, additional steps are required. To begin with, let us define the set of side frequencies for a given choice of pulse sequence and filter frequency $\omega_p$, $\Omega(\omega_p)=\{ m_1\omega_p,m_2\omega_p,m_3\omega_p,\ldots\}$. In general, the set contains an infinite number of frequencies, but it does not necessary include all possible multiples of $\omega_p$. For example, in the case of often considered Carr-Purcell \cite{Carr_PR1954} Meiboom-Gill \cite{Meiboom_RSI1958} (CPMG) pulse sequence the set is composed of only odd multiples because the Fourier coefficients of its filter function $c_m$ are non zero only when $m$ is an odd number \cite{\REVIEW}. It is also true that, in general, the coefficients $c_{m}$ go to zero for large $m$'s (e.g., in the case of CPMG sequence $c_{m}\propto m^{-1}$ \cite{\REVIEW}). The problem to solve here is to determine which frequency $\omega_p$ or $m\omega_p\in\Omega(\omega_p)$ is in-resonance with spectral line position $\omega_0$ so that $\delta\omega_\mathrm{min} = |\omega_0 - m\omega_p|\ll 2\pi/T_\mathrm{max}$. The most straightforward approach is to inspect another course of coherence evolution, but this time with the filter frequency set to be equal to the first side frequency of the original data set, i.e., to measure and plot $\{ W(T_1|m_1\omega_p),\ldots,W(T_\mathrm{max}|m_1\omega_p)\}$. If this plot belongs in the mismatched category, then, we can conclude that $m_1\omega_p$ is off-resonance, and we can check the next side frequency $m_2\omega_p$ by plotting $\{ W(T_1|m_2\omega_p),\ldots,W(T_\mathrm{max}|m_2\omega_p)\}$. If this plot also counts as mismatched, we proceed with the next side frequency, and so on. In practice, we only need to consider the few first elements of $\Omega(\omega_p)$ because of the tendency of Fourier coefficients to decrease with the multiplier $m$. Indeed, if we again take the example of the CPMG sequence and suppose that, say, $7\omega_p\approx \omega_0$, then the resonant term in the attenuation function would be given by $|c_{7}|^2|h_T(\delta\omega_\mathrm{min})|^2\sim T^2/7^2$ resulting in strongly suppressed polynomial scaling that is more likely to be included in the mismatched category on the first place. If all of the investigated side frequencies are mismatched, we conclude, by the process of elimination, that it is the original $\omega_p$ that is in-resonance. If, instead, we obtain the plot corresponding to one of the side frequencies, say, $m_k\omega_p$, that belongs in matching category, then we know that none of $\omega_p$, $m_1\omega_p,\ldots,m_{k-1}\omega_p$ were in-resonance, and the problem is redefined to determining whether $m_k\omega_p$ or one of {\it its} side frequencies $m m_k \omega_p \in \Omega(m_k\omega_p)=\{ m_1 m_k\omega_p, m_2 m_k \omega_p,\ldots\}$ is in-resonance---the problem that is solved with the same method as one described above.

\section{Numerical simulation of discrete spectrum spectroscopy}\label{sec:example}

Here, we present a numerical simulation of the implementation of the method for localization of spectral line positions described in Sec.~\ref{sec:disc_spectroscopy}.

The setup is identical to one used for the purpose of creating Figs.~\ref{fig:transitions} and \ref{fig:coherence}. The discrete spectral line positions have been chosen at random from probability distributions $p_k(\omega_k)\propto \exp[{-}(\omega_k-k\overline{\Delta\omega})^2/(2(0.02\overline{\Delta\omega}))^2]$ (the typical line separation $\overline{\Delta\omega}$ will be used as a unit of frequency) with the corresponding spectral line intensities,
\begin{align}
I(\omega_k) = \frac{2\nu^2\overline{\Delta\omega}\tau_c}{1+\tau_c^2\omega_k^2},
\end{align} 
where $\tau_c = 0.1\overline{\Delta\omega}$ and $\nu = \overline{\Delta\omega}$. We assume the CPMG pulse sequence, which results in Fourier coefficients \cite{Ajoy_PRA2011,\ACC,Krzywda_NJP2019},
\begin{align}
&c_m = \left\{\begin{array}{cl} \frac{2}{i\pi m}e^{im \frac{\pi}{2}}&\ \text{for $m\in$ Odd}\\[.1cm] 0 &\ \text{otherwise}\\\end{array}\right..
\end{align}
From Fig.~\ref{fig:transitions}, we can readout that the limit of course-grained regime is $T_\mathrm{trans}= 2\pi/\overline{\Delta\omega}\approx 6/\overline{\Delta\omega}$, i.e., beyond this duration length the attenuation function no longer scales linearly with $T$.

\begin{figure}[tb]
\centering
\includegraphics{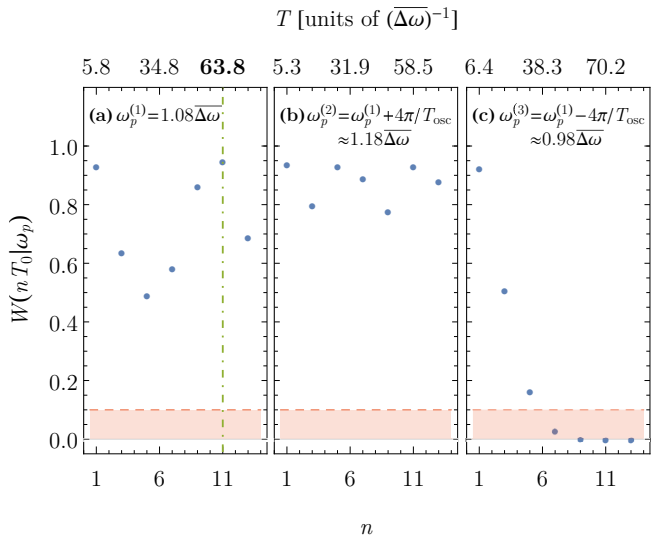}
\caption{The plots of coherence evolution data sets gathered for three subsequent choices of filter frequency. (a)~The first choice of $\omega_p=\omega_p^{(1)}$ is classified as mismatched because of visible oscillatory behavior. The vertical green dot-dashed line indicates the estimate of oscillation half-period $T_\mathrm{osc}/2=2\pi/\delta\omega_\mathrm{min}^{(1)}$, which then can be utilized to steer the next choices of filter frequency towards the spectral line position. (b)~The second choice $\omega_p = \omega_p^{(2)}=\omega_p^{(1)}+\delta\omega_\mathrm{min}^{(1)}=\omega_p^{(1)}+4\pi/T_\mathrm{osc}$ produced the plot classified in the mismatched category. (c)~The third choice $\omega_p=\omega_p^{(3)}=\omega_p^{(1)}-4\pi/T_\mathrm{osc}$ gives the matching plot.}
\label{fig:example}
\end{figure}

We start by selecting an arbitrary filter frequency $\omega_p = \omega_p^{(1)} = 1.08\,\overline{\Delta\omega}$, and collecting a data set composed of coherence evolving over increasing duration $\mathcal{W}^{(1)} = \{ W(T_0^{(1)}|\omega_p^{(1)}),\ldots,W(13 T_0^{(1)}|\omega_p^{(1)})\}$, where $T_0^{(1)}=2\pi/\omega_p^{(1)} = 5.8/\overline{\Delta\omega} \approx T_\mathrm{trans}$. Next, we plot the data in Fig.~\ref{fig:example} (a). Since we can clearly see oscillations of coherence, we classify it in the mismatched category and conclude that $\omega_p^{(1)}$ is off-resonance. Moreover, because we can clearly see a revival of coherence, we can estimate the detuning $\delta\omega_\mathrm{min}^{(1)}$ between $\omega_p^{(1)}$ and the nearest spectral line position $\omega_0$,
\begin{align}
\frac{63.8}{\overline{\Delta\omega}}\approx\frac{T_\mathrm{osc}}2 =\frac{2\pi}{\delta\omega_\mathrm{min}^{(1)}}\Rightarrow \delta\omega_\mathrm{min}^{(1)}=\frac{4\pi}{T_\mathrm{osc}}\approx 0.1\,\overline{\Delta\omega}.
\end{align}
Here, $T_\mathrm{osc}$ is the oscillation period of resonance term in the attenuation function that we introduced in Sec.~\ref{sec:sinc}.

We can now utilize the information on detuning to help us steer the choice of subsequent filter frequencies towards the line position. Since the detuning is a distance between the filter frequency and the line position $\delta\omega_\mathrm{min}^{(1)} = |\omega_0 - \omega_p^{(1)}|$, the data we have acquired so far is insufficient to determine whether $\omega_p^{(1)}$ is larger or smaller than $\omega_0$; the only way to find out is to check both options. Hence, as a second step, we set $\omega_p = \omega_p^{(2)}=\omega_p^{(1)}+\delta\omega_\mathrm{min}^{(1)}$ to be the next choice of filter frequency. The gathered data set $\mathcal{W}^{(2)}=\{ W(T_0^{(2)}|\omega_p^{(2)}),\ldots,W(13 T_0^{(2)}|\omega_p^{(2)})\}$ (with $T_0^{(2)}=2\pi/\omega_p^{(2)}\approx 5.3/\overline{\Delta\omega}$) is plotted in Fig.~\ref{fig:example} (b), and we see that it depicts a rapid oscillation with small amplitude---the features that make it belong in the mismatched category.

Thus, the next choice of filter frequency is $\omega_p=\omega_p^{(3)}=\omega_p^{(1)}-\delta\omega_{\mathrm{min}}^{(1)}$. The resultant data set $\mathcal{W}^{(3)}=\{ W(T_0^{(3)}|\omega_p^{(3)}),\ldots,W(13T_0^{(3)}|\omega_p^{(3)})\}$ with $T_0^{(3)}=2\pi/\omega_p^{(3)}\approx 6.4/\overline{\Delta\omega}$ is plotted in Fig.~\ref{fig:example} (c), and it depicts the monotonic decay of coherence that we classify in the matched category. Therefore, we conclude that frequency $\omega_0^{(\mathrm{est})}=\omega_p^{(3)}\approx 0.9847\,\overline{\Delta\omega}$ belongs in the discrete distribution of spectral line positions. The accuracy of this estimation is very good as the actual line position is $\omega_0^{(\mathrm{real})} \approx 0.9826\, \overline{\Delta\omega}$ so that the error is $|\omega_0^{(\mathrm{est})}-\omega_0^{(\mathrm{real})}|/\omega_0^{(\mathrm{real})} \approx 0.2$\%.

Note that when we used the detuning $\delta\omega_\mathrm{min}^{(1)}$ to steer the filter frequency we implicitly assumed that it described the distance between spectral line and the primary frequency $\omega_p^{(1)}$, as opposed to one of its side frequencies $m\omega_p^{(1)}\in\Omega(\omega_p^{(1)})$. Since $\omega_p^{(3)} = \omega_p^{(1)}-\delta\omega_\mathrm{min}^{(1)}$ proved to be in resonance this hypothesis has been confirmed. Indeed, in an event when $\delta\omega_\mathrm{min}^{(1)} = |\omega_0 - m\omega_p^{(1)}|$ neither $\omega_p^{(1)}+\delta\omega_\mathrm{min}^{(1)}$ nor $\omega_p^{(1)}-\delta\omega_\mathrm{min}^{(1)}$ would produce matching plot, which would indicate that the assumption was invalid. In such an event, we would have to continue our search by going over side frequencies and constructing plots for subsequent choices of filter frequencies of the form $\omega_p = m_k\omega_p^{(1)}\pm\delta\omega_\mathrm{min}^{(1)}$ until we find one that can be classified as matching.

\begin{figure}[tb]
\centering
\includegraphics{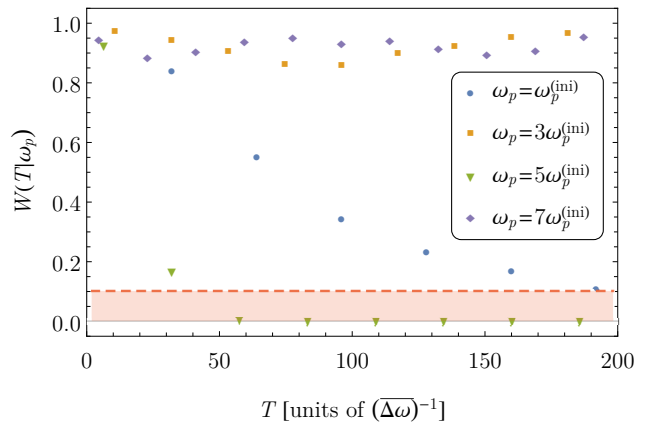}
\caption{The plots of coherence evolution data sets gathered for the initial choice of filter frequency $\omega_p=\omega_p^{(\mathrm{ini})}$ and its first three side frequencies: $\omega_p=3\omega_p^{(\mathrm{ini})}, 5\omega_p^{(\mathrm{ini})}, 7\omega_p^{(\mathrm{ini})}$. Only the initial (blue circles) and the third (green triangles) plot, corresponding to $\omega_p=\omega_p^{(\mathrm{ini})}$ and $\omega_p = 5\omega_p^{(\mathrm{ini})}$, respectively, are classified in the matching category, indicating that the resonance occurs for frequency $5\omega_p^{(\mathrm{ini})}$.
}
\label{fig:finding_m}
\end{figure}

Finally, we present an example when the matching plot is obtained for the initial choice of filter frequency $\omega_p = \omega_p^{(\mathrm{ini})} =0.1965\,\overline{\Delta\omega}$. In such an event, we have to take additional steps to determine if the in-resonance frequency is the primary $\omega_p^{(\mathrm{ini})}$ or one of its side frequencies $m\omega_p^{(\mathrm{ini})}\in\Omega(\omega_p^{(\mathrm{ini})})=\{ 3\omega_p^{(\mathrm{ini})},5\omega_p^{(\mathrm{ini})},\ldots\}$. The initial plot of coherence evolution is shown in Fig.~\ref{fig:finding_m}, and it exhibits a monotonic decay which qualifies for the matching category. Next, we plot the data set obtained for first side frequency $\omega_p = 3\omega_p^{(\mathrm{ini})}$, and we find it belongs in the mismatched category. So far, we have not disproved the hypothesis that $\omega_p^{(\mathrm{ini})}$ is in-resonance, and thus, we proceed with the next side frequency $\omega_p = 5\omega_p^{(\mathrm{ini})}$. The plot of the corresponding data set also exhibits monotonic decay that places it firmly in the matching category. Therefore, we conclude that the resonance occurred not for initial $\omega_p^{(\mathrm{ini})}$ but rather for $5\omega_p^{(\mathrm{ini})}\approx 0.9826\,\overline{\Delta\omega}=\omega_0^{(\mathrm{real})}$.

Repeating the procedure described above with different choices of the initial filter frequency allows to pinpoint more spectral lines.

\section{Conclusions}\label{sec:end}

We have analyzed the $T$-scaling laws (i.e., the parametric dependence of the attenuation function on the duration of the evolution) for discrete spectral density in single qubit dynamical decoupling spectroscopy. In this sense, the results presented in this paper can be considered as an extension to Ref.~\cite{\ACC}, where the scaling laws have been investigated only for continuous spectral densities. 

We have shown how these $T$-scaling laws can be utilized in a variant of spectroscopy that operates in the post-coarse-grained regime, i.e., for the range of duration lengths when the pulse sequence-induced frequency filters have high enough resolution to distinguish discrete spectral lines. The resultant discrete distribution of spectral line positions provides a direct glimpse into the structure of the energy spectrum of the environmental Hamiltonian. 

As a closing remark, we would like to point out what this paper is {\it not} about. We focused here on the case where the discrete spectral density emerges from its continuous approximation when the resolution of the frequency filters is high enough. This implies that spectra we considered composed of a large number of relatively densely packed and comparably intense spectral lines (see Appendix \ref{sec:cont_limit}). Therefore, we have {\it not} considered the cases of sparse discrete line distributions \cite{Childress_Science06,Zhao_NN12,Taminiau_PRL2012,Krzywda_PRA2017,Hernandez_PRB2018} nor a sparse discrete distribution superposed with a continuous background \cite{Zhao_PRL2011,Kolkowitz_PRL2012}. Nevertheless, our findings regarding the $T$-scaling laws could be applied in those cases as well. In particular, for the latter type of spectral density, the method for `fractional distillation' of the attenuation function into components with different $T$-scalings provide an effective tool for separating the background contribution from the discrete spectral lines. It is the design of post-coarse-grained spectroscopy described in Sec.~\ref{sec:disc_spectroscopy} that would not necessarily be the optimal choice for the information acquisition scheme for those types of spectra. Indeed, when one expects that the discrete line is isolated, the estimation of the period of the oscillation of the resonant passband (which is determined by the detuning) might already yield results with a satisfactory accuracy. In fact, the matching--mismatching dichotomy approach used here might be outright invalid in certain instances, because of the inescapable exponential decay of the coherence due to the linearly scaling continuous background. As a side note, recently \cite{Sakuldee_Spectroscopy_arXive2019}, it has been proposed that the problem with the background contribution could be circumvented with a mode of spectroscopy where, instead of dynamical decoupling control, the frequency filters are obtained through the sequence of projective measurements preformed on the qubit probe. The idle intervals between the measurement and the subsequent reinitialization of the probe present in the scheme, introduces a new feature to the resultant filters, that decouple the qubit from the continuous portion of spectral density.

Moreover, we have carried out our investigation under the assumption of Gaussian approximation, so our results cannot be directly compared with more specific non-Gaussian models that also involve some form of discrete frequency distributions, such as Ref.~\cite{Kotler_PRL2013}. Nevertheless, we speculate that the method of matching--mismatching dichotomy as a mean to localize spectral line positions, could still be applied even beyond Gaussian approximation. This speculation is partially supported by the results found in Refs.~\cite{Cywinski_PRB2009,Malinowski_NN2017,Malinowski_PRL2017} and by the following qualitative argument. It is expected that the non-Gaussian terms in attenuation function (i.e., the cumulants of order higher than two) are bounded by the second-order, purely Gaussian term \cite{\REVIEW}. Hence, when the filter passbands are detuned from spectral lines and the Gaussian contribution drops, the non-Gaussian terms would be suppressed. On the other hand, when the passband is in resonance and the second-order terms blows up, the non-Gaussian terms would not interfere with it.

\section*{Acknowledgments}
This work was supported by funds of Polish National Science Center (NCN), grant no.~2015/19/B/ST3/03152.

P.~S. would like to thank \L{}.Cywi\'{n}ski for inspiring discussions and many helpful comments on the paper.

\appendix

\section{Fourier transform of time-domain filter function $f_{\omega_p}(t)$}\label{sec:filter}

Consider a `windowed' function $G_{[0,T]}(t)$ that is identically zero outside the interval $[0,T]$. Like any function defined on a interval, $G_{[0,T]}$ can be decomposed in the discrete Fourier basis $\{ e^{-i l \frac{2\pi}{T} t} \}_{l\in \mathrm{Integers}}$,
\begin{align}
G_{[0,T]}(t) = \Theta(T-t)\Theta(t)\sum_{l = -\infty}^\infty s_l(T) e^{i l(2\pi/T) t}
\end{align}
where the expansion coefficients are given by
\begin{align}
s_l(T) = \frac{1}{T}\int_0^T G_{[0,T]}(t) e^{-i l (2\pi/T) t}dt.
\end{align}

Now, consider a periodic function,
\begin{align}
F_{T_0}(t) = F_{T_0}(t - k T_0),
\end{align}
which, according to Fourier theorem, can be decomposed into a series,
\begin{align}
F_{T_0}(t) &= \sum_{m = -\infty}^\infty c_m e^{i m (2\pi/T_0) t}
= \sum_{m = -\infty}^\infty c_m e^{i m \omega_p t}
\end{align}
where we have defined $\omega_p = 2\pi / T_0$ and the Fourier series coefficients are computed with respect to the period $T_0$,
\begin{align}
c_m = \frac{1}{T_0}\int_{0}^{T_0} F_{T_0}(t) e^{-i m \omega_p t} dt.
\end{align}
Let us set the window width to be equal to some integer multiple of the period $T = n T_0$ and use the periodic function to construct our windowed $G$,
\begin{align}
G_{[0,n T_0]}(t) = \Theta(n T_0 - t)\Theta(t)F_{T_0}(t).
\end{align}
If we recalculate Fourier coefficients $s_l(n T_0)$, we will discover that they no longer depend on the length of duration $T$ (more precisely, on the multiplier $n$),
\begin{align}
\nonumber
s_{l}(n T_0) & = \frac{1}{n T_0}\int_0^{n T_0} G_{[0,n T_0]}(t) e^{-i l (2\pi/n T_0) t}dt \\
\nonumber
& = \frac{1}{n T_0}\sum_{r = 0}^{n-1} \int_{r T_0}^{(r+1)T_0} F_{T_0}(t) e^{-i l (2\pi/n T_0) t} dt \\
\nonumber
& = \frac{1}{n T_0}\sum_{r = 0}^{n-1} \int_0^{T_0} F_{T_0}(t - r T_0)e^{-i l (2\pi/n T_0) t } e^{i l 2\pi(r/n)}dt\\
\nonumber
& = \left(\frac{1}{n}\sum_{r = 0}^{n-1}e^{i l 2\pi(r/n)}\right)\frac{1}{T_0}\int_0^{T_0} F_{T_0}(t)e^{-i l (2\pi/n T_0)t}dt\\
&= \delta_{l,m n}\frac{1}{T_0}\int_0^{T_0} F_{T_0}(t)e^{-i l (2\pi/n T_0)t}dt = \delta_{l,m n} c_m.
\end{align} 

In our case, the widowed function is, of course, the time-domain filter function $G_{[0, nT_0]}(t) = f_{\omega_p}(t)$. Hence, to calculate its Fourier transform, we can utilize the corresponding Fourier series decomposition,
\begin{align}
\nonumber
\tilde{f}_{\omega_p,n T_0}(\omega) = \int_{-\infty}^\infty\!\!  \Theta(nT_0-t)\Theta(t)\!\! \sum_{m = -\infty}^\infty c_m e^{i m \omega_p t} dt&\\
\nonumber
= \sum_{m = -\infty}^\infty c_m \int_0^{nT_0} e^{i (m \omega_p - \omega) t} dt&\\
\nonumber
= \sum_{m=-\infty}^\infty c_m \frac{e^{i (m\omega_p- \omega) nT_0} - 1}{i (m \omega_p - \omega)}&\\
= \sum_{m=-\infty}^\infty c_m n T_0 e^{-i (nT_0\omega/2)}\mathrm{sinc}\left(\frac{n T_0(\omega - m \omega_p)}{2}\right).&
\end{align}

\section{Coarse-grained approximation to discrete spectra}\label{sec:cont_limit}

To show how the attenuation function for the discrete spectrum can be approximated by the integral over the continuous spectral density, we start by rewriting Eq.~\eqref{eq:disc_chi} as
\begin{align}
\nonumber
\chi(T|\omega_p) &=\sum_{k=1}^N I(\omega_k)T^2F(T\omega_k)=\sum_{k=1}^N s_k T^2F(T\omega_k) \Delta\omega_k,
\end{align}
where $\omega_1<\omega_2<\ldots<\omega_N$, $\Delta\omega_k=\omega_{k+1}-\omega_k$, $s_k = I(\omega_k)/\Delta\omega_k$, and
\begin{align}
\nonumber
&T^2F(T\omega) = |\tilde{f}_{\omega_p,T}(\omega)|^2\\
\nonumber
&=\sum_{m,m'}c_{m}c_{m'}^*h_T(\omega-m\omega_p)h_T(\omega-m'\omega_p)\\
\nonumber
&=T^2 \sum_{m}c_m \mathrm{sinc}\left[\frac{T\omega-Tm\omega_p}2\right]\\
&\phantom{=}\times\sum_{m'}c_{m'}^* \mathrm{sinc}\left[\frac{T\omega-Tm'\omega_p}2\right].
\end{align}
Now, consider a smooth function $\bar{S}(\omega)$ such that
\begin{enumerate}
\item $\bar{S}(\omega_k) = s_k$ for each $\omega_k$,
\item $\min(s_k,s_{k+1})\leqslant \bar{S}(\omega) \leqslant \max(s_k,s_{k+1})$ for each interval $\omega\in [\omega_k,\omega_{k+1}]$.
\end{enumerate}
Essentially, function $\bar{S}(\omega)$ is an interpolation of discrete distribution of spectral line intensity $I(\omega_k)$, which we can use to define a coarse-grained version of attenuation function,
\begin{align}
\nonumber
\bar{\chi}(T|\omega_p)&\equiv \int_{-\infty}^\infty  \bar{S}(\omega)T^2 F(T\omega)d\omega\\
&=\sum_k \int_{\omega_k}^{\omega_{k+1}} \bar{S}(\omega) T^2 F(T\omega)d\omega
\end{align}
In the limit where the width of frequency-domain filter $F$ is such that $T\Delta\omega_k\ll 1$ for each $\Delta\omega_k$ so that we have
\begin{align}
&T^2 F(T\omega_k) \approx T^2 F(T\omega_{k+1}),
\end{align}
we get that each term in the above sum satisfies
\begin{align}
\nonumber
T^2 F(&T\omega_k)\min(s_k,s_{k+1})\Delta\omega_k\\
\nonumber
&\leqslant \int_{\omega_k}^{\omega_{k+1}}\!\!\! \!\!\! T^2 F(T\omega)\bar{S}(\omega)d\omega \\
&\leqslant T^2 F(T\omega_k)\max(s_k,s_{k+1})\Delta \omega_k.
\end{align}
In this limit the `sandwiching' expressions only differ by
\begin{align}
\nonumber
|T^2 F(&T\omega_k)s_k \Delta\omega_k -T^2 F(T\omega_k)s_{k+1}\Delta\omega_k |\\
\nonumber
&\leqslant T^2 \Delta\omega_k \left|s_k-s_{k+1}\right|\\
&\equiv T^2 \Delta\omega_k\Delta s_k,
\end{align}
and when, in addition to $T\Delta\omega_k\ll 1$, we also have that the variation of line intensity on each interval $\Delta\omega_k$ is not too large $T^2\Delta\omega_k\Delta s_k \ll 1$, we get that the fine details of the interpolation scheme [i.e., the course of $\bar{S}(\omega)$ in between line positions] do not impact the overall value of the integral and that
\begin{align}
\bar{\chi}(T|\omega_p)\approx \sum_k T^2 F(T\omega_k)s_k\Delta\omega_k = \chi(T|\omega_p).
\end{align}
When this is the case, $\bar{S}(\omega)$ is identified with the coarse-grained spectral density $S(\omega)$.

\bibliography{cont_disc_trans}
\end{document}